\documentclass{article}
\usepackage[margin=3cm]{geometry}
\usepackage{authblk}

\usepackage{amsmath,amssymb,amsfonts}
\DeclareMathOperator\arctanh{arctanh}

\usepackage{mathtools}
\usepackage{physics} 

\usepackage{amsthm}
\theoremstyle{plain}
\newtheorem*{theorem}{Theorem}

\usepackage{todonotes}

\usepackage[nocompress]{cite}

\usepackage{caption}
\usepackage{subcaption}

\usepackage{graphicx}

\renewcommand{\dd}{\mathrm{d}}

\numberwithin{equation}{section}

\usepackage{hyperref}
\hypersetup{colorlinks=true,allcolors=blue,urlcolor=cyan}

\title{Azimuthal geodesics in closed FLRW cosmological models}

\author[1,2]{Christian G B\"ohmer\footnote{Email: c.boehmer@ucl.ac.uk}}
\author[1]{~Antonio d'Alfonso del Sordo\footnote{Email: a.dalfonsodelsordo@ucl.ac.uk}}

\author[1]{~Betti Hartmann\footnote{Email: b.hartmann@ucl.ac.uk}}
\author[3]{Linus Elias M\"unch\footnote{Email: eliasmue@uni-bonn.de}}
\affil[1]{Department of Mathematics, University College London, \authorcr Gower Street, London WC1E 6BT, UK\medskip}
\affil[2]{Astrophysics Research Centre, School of Mathematics, \authorcr Statistics and Computer Science, University of KwaZulu-Natal, \authorcr Private Bag X54001, Durban 4000, South Africa\medskip}
\affil[3]{Rheinische Friedrich-Wilhelms-Universit\"at Bonn, 53012 Bonn, Germany}

\date{\today}

\begin{document}

\maketitle

\begin{abstract}
    We study geodesics in Friedmann-Lema{\^\i}tre-Robertson-Walker (FLRW) cosmological models and give the full set of solutions. For azimuthal geodesics, in a closed universe, we give the angular distance travelled by a test particle moving along such a geodesic during one cycle of expansion and re-collapse of the universe. We extend previous results regarding the path followed by light rays to the two-fluid case, also including a cosmological constant, as well as to massive test particles. Our work contains various new results and explicit formulae, often using special functions which naturally appear in this setting.
\end{abstract}

\clearpage 

\tableofcontents

\clearpage

\section{Introduction}

The Friedmann-Lema\^itre-Robertson-Walker (FLRW) spacetimes are four-dimensional models satisfying the cosmological principle, which states that the universe is homogeneous and isotropic. 
They are solutions to the Einstein field equations of General Relativity (GR) with energy-momentum content given by perfect, barotropic fluids, with linear equation of state $p=w\rho$, for a constant $w$. 
They provide a robust mathematical framework for modelling the large-scale structure of our expanding universe\cite{SupernovaSearchTeam:1998fmf,SupernovaCosmologyProject:1998vns,Wright:1992tf,WMAP:2006bqn,Planck:2019evm}. 
Named after its contributors~\cite{Friedman:1922kd,Lemaitre:1927zz,Robertson:1935zz,Robertson:1935jpx,Robertson:1936zz,Walker:1937qxv}, the FLRW metric describes a maximally symmetric space of constant curvature evolving in size. The curvature of the spatial surfaces dictates the geometry of the universe on cosmic scales~\cite{Lachieze-Rey:1995qrb,Hawking:1973uf,Griffiths:2009dfa}, and the size is governed by a \emph{scale factor} which appears in the metric.

The spatial curvature can take one of three forms: negative, zero, or positive, corresponding to an open, flat, or closed universe, respectively. The geometry of a closed universe, which is spatially a sphere, allows for a scenario in which the universe expands up to a maximum size, before recollapsing during the so-called \emph{big crunch}. This oscillatory behaviour was first noted by Friedmann~\cite{Friedman:1922kd}. In the realm of cosmology, the concept of oscillating universes has been a subject of theoretical exploration. Notably, in~\cite{Harrison:1967zz}, Harrison provided a classification for FLRW models, including oscillatory universes. These models of a universe undergoing cycles of expansion and contraction were also further explored in~\cite{Barrow:1995cfa,Barrow:2004ad,Brandenberger:2016vhg,Barrow:2017zar}.
In particular, in~\cite{Barrow:1986cur}, the closed-universe recollapse conjecture---which states that, under certain conditions, the expanding universe could eventually reverse its course, contracting back to a highly dense state, and initiating a new cycle of expansion and contraction---was proposed as a potential solution to cosmological issues, such as the \emph{singularity problem}. One such model, based on the \emph{Ekpyrotic model} \cite{Khoury:2001wf}, was introduced in \cite{Steinhardt:2004gk}. This model contains two four-dimensional branes in a higher dimensional space. All matter is confined to the brane, while gravity as a property of  itself acts in all dimensions. The Big Bang corresponds to the collision of these two branes with inter-brane distance given by a scalar field. The potential associated to this scalar field leads to 
repeated collisions and successive bounces of the branes. Dark energy can be reinterpreted as a small attractive force between branes within this model \cite{Lehners:2008vx}. Cyclic models are also considered as alternatives to cosmic inflation \cite{Linde:2002ws}. In \cite{Pavlovic:2020sei}, model-independent properties of cyclic cosmologies were explored and, subsequently, applied to modifications of gravity, such as $f(R)$ and $f(T)$-theories.
To obtain information regarding the scale factor, one can solve the Einstein field equations in the context of the FLRW metric. Some of the solutions to the cosmological field equations can be found in terms of elementary functions, see e.g.~\cite{Bohmer:2016ome,Baumann:2022mni}. However, in some cases (for example, involving terms such a \emph{cosmological constant}, non-zero curvature, and a cosmological fluid) explicit solutions to the cosmological field equations have been obtained in terms of \emph{special functions}, often encountered in problems within mathematical physics~\cite{andrews1999special,armitage2006elliptic,Halburd2017special}, such as elliptic functions~\cite{Edwards:1972,Edwards:1973,Coquereaux:1981ya,DAmbroise:2009ruz,Chen:2014fqa,Faraoni:2021opj,Pavlov:2021vfv}.

While the prevailing observational evidence has strongly favoured a spatially flat universe over a closed one, recent observations have prompted a reconsideration of the possibility that the universe may in fact exhibit a closed geometry~\cite{DiValentino:2019qzk,Handley:2019tkm,Vagnozzi:2020rcz,Vagnozzi:2020dfn,DiValentino:2020hov,Dhawan:2021mel,Anselmi:2022uvj,Glanville:2022xes,Semenaite:2022unt}.The 2018 PLANCK data showed that
$k > 0$ at 3.4$\sigma$ \cite{Planck:2018vyg}. 
It hence makes sense to consider test particle motion also in closed FLRW spacetimes. In this geometric setting, it is possible to study azimuthal geodesics.
In GR, azimuthal geodesics have been extensively studied in the context of rotating systems and settings involving spherical symmetry, for example the movement around massive objects such as black holes, e.g.\cite{Chandrasekhar:1985kt}, where special functions arise~\cite{Lammerzahl:2015qps,Cieslik:2022uki}. Azimuthal geodesics within closed FLRW models were discussed in our recent work~\cite{Boehmer:2024kdt}, which specifically focused on azimuthal null geodesics. We found that the angular distance travelled by a ray of light starting at the beginning of the universe during the expansion and recollapse is given by $\Delta\varphi=2\pi/(1+3w)$, for an arbitrary linear equation of state parameter $w$. We now broaden the scope to encompass several other cases. 

In particular, we begin our discussion, in Sec.~\ref{sec:fieldequations}, by introducing the geodesic equations for FLRW in terms of the conserved quantities arising from the symmetry of the spacetime. In Sec.~\ref{section:solutions}, we provide a complete set of solutions to the geodesic equations for the angular and radial motions. Section~\ref{sec:azimuthalgeodesics} deals with azimuthal geodesics as well as the general formula for the angular distance travelled by a geodesic test particle in a closed FLRW spacetime. We first focus on the azimuthal distance travelled by a massless test particle in two-fluid cosmologies extending previous results, in  Sec.~\ref{sec:masslesstwofluid}. In Sec.~\ref{sec:cosmologicalconst}, we consider the case in which one of the two fluids represents a cosmological constant. Einstein first introduced a positive cosmological constant in his field equations to achieve a static universe~\cite{Einstein:1917ce}, and it is now a potential candidate for \emph{dark energy}, responsible for the accelerated expansion of the universe, for a review see \cite{Copeland:2006wr}.  In Sec.~\ref{sec:massiveone}, we turn to massive test particles in single fluid cosmologies. In most cases, these results have to be expressed in terms of special functions, in particular, elliptic and hypergeometric functions. 

\paragraph{Notation.} In the following, we will set the speed of light $c$ and Newton's constant $G$ to unity. 

\section{Field equations and geodesics} \label{sec:fieldequations}
\subsection{Field equations}

In the following, we are interested in the geodesic motion in a Friedmann-Lema\^\i tre-Robertson-Walker (FLRW) spacetime with line element
\begin{equation}
    \dd s^2=-\dd t^2+a^2(t)\left[\frac{\dd r^2}{1-kr^2}+r^2\dd\theta^2+r^2\sin^2\negmedspace\theta \dd\varphi^2\right] ,
    \label{eq:FLRWline}
\end{equation}
where $k$ can take values $\{-1,0,1\}$. 

A constant time slice of this spacetime corresponds to a maximally symmetric three-dimensional space with constant positive ($k=1$), zero ($k=0$) or negative ($k=-1$) curvature. 

When $k=1$, the so-called scale factor $a(t)$ evolves according to the Friedmann equations
\begin{align}
    \left(\frac{\dot{a}}{a}\right)^2&=\frac{8\pi}{3}\left(\rho_1+\rho_2\right)-\frac{1}{a^2}\,, \label{eq:friedmanna}\\
    \frac{\ddot{a}}{a}&=-\frac{4\pi}{3}\left[\left(\rho_1+3p_1\right)+\left(\rho_2+3p_2\right)\right],
    \label{eq:friedmannb}
\end{align}
where the dot denotes the derivative with respect to cosmic time $t$. The evolution of $a(t)$ is determined by the energy-momentum content and we assume the universe to be filled with two perfect, barotropic fluids with equation of state $p_i(t)=w_i \rho_i(t)$, for $i\in\{1,2\}$, where $\rho_i(t)$ is the energy density, $p_i(t)$ is the pressure, and $w_i$ is the equation of state parameter. Equations (\ref{eq:friedmanna}) and (\ref{eq:friedmannb}) imply the energy-momentum conservation, 
\begin{equation}
    \dot{\rho}_i+3\frac{\dot{a}}{a}(\rho_i+p_i)=0 \,, \quad \text{for}\; i\in\{1,2
  \}\,,
    \label{eq:conservation}
\end{equation}
which is satisfied by each fluid component, and yields
\begin{equation}
  \rho_i=\rho_i^{(0)} a^{-3(1+w_i)} \,, \quad 
  \text{for}\; i\in\{1,2\}\,,
  \label{eq:rhoanda}
\end{equation}
where $\rho_i^{(0)}$, for $i\in\{1,2\}$, is a constant of integration. We now define $\beta_i\coloneqq8\pi\rho_i^{(0)}/3$, and $\gamma_i\coloneqq1+3w_i$, for $i\in\{1,2\}$, and re-write Eq.~\eqref{eq:friedmanna} as
\begin{equation}
    \left(\dot{a}\right)^2=\beta_1 a^{-\gamma_1}+\beta_2 a^{-\gamma_2}-1 \,.
    \label{eq:friedmann1a}
\end{equation}

\subsection{Geodesic equations}

The geodesic equation is given by
\begin{equation}
\label{eq:geo}
    \frac{{\rm d}^2{x}^{\mu}}{{\rm d}\lambda^2} + \Gamma^{\mu}_{\alpha\beta} \frac{{\rm d}{x}^{\alpha}}{{\rm d}\lambda} \frac{{\rm d}{x}^{\beta}}{{\rm d}\lambda}=0\,,
\end{equation}
and can be derived by considering the Lagrangian $\mathcal{L}$
\begin{equation}
    \mathcal{L}=g_{\mu\nu}\frac{\dd x^{\mu}}{\dd \lambda}\frac{\dd x^{\nu}}{\dd \lambda}\,,
\end{equation}
with affine parameter $\lambda$. Note that $\mathcal{L}=0$ for null geodesics 
and $\mathcal{L}=-1$ for timelike geodesics, respectively. For the metric (\ref{eq:FLRWline}), we have the Lagrangian
\begin{equation}
    \mathcal{L}= -{t'}^2+a^2(t)\left[\frac{{r'}^2}{1-k r^2} + 
    r^2{\theta'}^2 + r^2\sin^2\negmedspace\theta {\varphi'}^2\right],
    \label{eq:azi1}
\end{equation}
with the prime denoting derivatives with respect to $\lambda$. The four components of the geodesic equation~(\ref{eq:geo}) are
\begin{align}
    {t''}+a \dot{a}\left(\frac{{r'}^{2}}{1-k r^{2}}+r^{2} {\theta'}^{2}+r^{2} \sin ^{2} \theta {\varphi'}^{2}\right) &= 0 \,,
    \label{eq:geo1}\\
    {r''} + 2 \frac{\dot{a}}{a} {t'} {r'}+\frac{k r}{1-k r^{2}} {r'}^{2}-
    r\left(1-k r^{2}\right) \left({\theta'}^{2}+\sin ^{2} \theta {\varphi'}^{2}\right) &= 0 \,,
    \label{eq:geo2}\\
    {\theta''}+ 2 \frac{\dot{a}}{a} {t'} {\theta'}+\frac{2}{r} {r'} {\theta'}-
    \cos \theta \sin \theta {\varphi'}^{2} &= 0 
    \label{eq:geo3} \,, \\
    {\varphi''}+2 \frac{\dot{a}}{a} {t'} {\varphi'}+\frac{2}{r} {r'} {\varphi'}+2 \frac{\cos \theta}{\sin \theta} {\varphi'} {\theta'} &=0 \,.
    \label{eq:geo4}
\end{align}
Due to the six isometries of the FLRW spacetime, we can define conserved quantities associated with those.

\subsection{Conserved quantities}
There exist six Killing vectors for the metric~(\ref{eq:FLRWline}) that satisfy the Killing equations $\nabla_{\mu}\xi_{\nu} + \nabla_{\nu} \xi_{\mu}=0$. Explicitly, these are given by
\begin{eqnarray*}
    \xi_{(1),\mu}&=& a^2
    \left(0, 
    \frac{1}{\sqrt{1-kr^2}}\sin{\theta}\cos{\varphi}, 
    r\sqrt{1-kr^2}\cos{\theta}\cos{\varphi},
    -r\sqrt{1-kr^2}\sin\theta\sin{\varphi}\right), \\
    \xi_{(2),\mu}&=& a^2
    \left(0,
    \frac{1}{\sqrt{1-kr^2}}\sin{\theta}\sin{\varphi},
    r\sqrt{1-kr^2}\cos{\theta}\sin{\varphi},
    r\sqrt{1-kr^2}\sin\theta\cos{\varphi}\right), \\
    \xi_{(3),\mu}&=& a^2
    \left(0,
    \frac{1}{\sqrt{1-kr^2}}\cos{\theta},
    -r\sqrt{1-kr^2}\sin{\theta},
    0\right), \nonumber \\
    \xi_{(4),\mu} &=& a^2
    \left(0,0,
   r^2\sin\varphi,
    r^2\cos\theta\sin\theta\cos\varphi\right), \\
    \xi_{(5),\mu} &=& a^2
    \left(0,0,
  -r^2\cos\varphi,
    r^2\cos\theta\sin\theta\sin\varphi\right), \\
    \xi_{(6),\mu} &=& a^2
    \left(0,
    0,
   0,
    -r^2 \sin^2\theta\right).
\end{eqnarray*}
In Euclidean space, these correspond to three translations and three rotations and are thus related to the conserved linear and angular momentum, respectively. The existence of Killing vectors implies the existence of conserved quantities $\mathcal{C}_{(i)} := \xi_{(i),\mu} x'^{\mu}(\lambda)$ which are constant along the geodesics, that is, 
\begin{align}
   \frac{a^2}{\sqrt{1-kr^2}}\sin{\theta}\cos{\varphi}r'+a^2 r\sqrt{1-kr^2}\cos{\theta}\cos{\varphi}\theta'-a^2r\sqrt{1-kr^2}\sin\theta\sin{\varphi}\varphi'&=p_x\,, \label{eqn:px}\\
   \frac{a^2}{\sqrt{1-kr^2}} \sin{\theta}\sin{\varphi}r'+a^2r \sqrt{1-kr^2}\cos{\theta}\sin{\varphi}\theta'+a^2r\sqrt{1-kr^2}\sin\theta\cos{\varphi}\varphi'&=p_y\,, \label{eqn:py}\\
   \frac{a^2}{\sqrt{1-kr^2}}\cos\theta r'-a^2r\sqrt{1-kr^2}
   \sin\theta \theta'&=p_z\,, \label{eqn:pz}\\
   a^2r^2\sin\varphi\theta'+a^2r^2\sin^2\theta\cot\theta\cos\varphi\varphi'&=L_x\,, \label{eqn:lx}\\
   -a^2r^2\cos{\varphi}\theta'+a^2r^2\sin^2\theta\cot\theta\sin\varphi\varphi'&=L_y\,, \label{eqn:ly}\\
   -a^2r^2\sin^2{\theta}\varphi'&=L_z\,. \label{eq:geophi}
\end{align}
However, these are not all independent, only $p_z$, $L_z$, $p^2=p_x^2 + p_y^2 + p_z^2$ and $L^2=L_x^2 + L_y^2 + L_z^2$ are independent. A direct calculation shows that $L^2$ and $p^2$ take the neat forms
\begin{equation}
    L^2=a^4 r^4 \left(\theta'\right)^2+\csc^2\theta L_z^2\,,
    \qquad
    p^2=\frac{a^4 \left(r^2 \left(1-k r^2\right)^2 \left(\left(\theta '\right)^2+\sin ^2(\theta
   ) \left(\varphi '\right)^2\right)+\left(r'\right)^2\right)}{1-k r^2} \,. 
    \label{eq:cons}
\end{equation}
It is convenient to rewrite these equations as follows
\begin{align}
    (\theta')^2 &= \frac{L^2}{a^4 r^4}-\frac{L_z^2}{a^4 r^4\sin^2\theta}\,, \label{eq:geotheta} \\
    (r')^2 &= \frac{1-kr^2}{a^4}\left(p^2+kL^2-\frac{L^2}{r^2}\right).
    \label{eq:geor}
\end{align}
In this form, it is clear that one is dealing with two coupled first order differential equations. Note that, from the $\theta$-equation \eqref{eq:geotheta}, we obtain a restriction for the $\theta$-motion, that is,
\begin{equation}
    \arcsin\left(\frac{L_z}{L}\right)\leq\theta\leq \pi-\arcsin\left(\frac{L_z}{L}\right).
\end{equation}
When $L_z=L$, the motion ensues in the equatorial plane, $\theta=\pi/2$.
Finally, the $t$-component of the geodesic equation becomes
\begin{equation}
    t''+(kL^2+p^2)\frac{\dot{a}}{a^3}=0\implies
\label{eq:geot}
    (t')^2-\frac{(kL^2+p^2)}{a^2}=\mathcal{L},
\end{equation}
where $\mathcal{L}=0$ for massless test particles and $\mathcal{L}=-1$ for massive test particles, respectively. 
We will present solutions to the equations (\ref{eq:geophi}), (\ref{eq:geotheta}), (\ref{eq:geor}) and (\ref{eq:geot}) in Section \ref{section:solutions}. 

\subsection{Azimuthal geodesics}

As we are primarily interested in the study of azimuthal geodesics in a closed universe ($k=1$), let us investigate the full geodesic equation assuming $r=r_0=\text{constant}$. This means $r''=r'=0$ which together with~(\ref{eq:geo2}) implies $(1-r_0)=0$ or $r_0=1$ since the other possible solution $r_0=0$ or both $\theta$ and $\varphi$ constant are of no interest to us. Hence, $r_0=1$ is the only possible solution for such geodesics.

When $r_0=1$ is used in~(\ref{eq:cons}), it implies $p=0$ and, together with~(\ref{eqn:px})--(\ref{eqn:pz}), yields $p_x=p_y=p_z=0$. Consequently, Eq.~(\ref{eq:geor}) is also identically satisfied.

Next, we assume $\theta=\theta_0=\text{constant}$. When this is put into~(\ref{eq:geo3}), one immediately arrives at $\cos\theta_0\sin\theta_0{\varphi'}^2=0$. Since solutions with constant $\varphi$ are of no interest to us, we are left with $\theta_0=\pi/2$. We neglect the solution $\theta_0=0$ as it corresponds to a pole. This yields $L_x=L_y=0$ by~(\ref{eqn:lx}) and~(\ref{eqn:ly}), and $L=L_z=-a^2\varphi'$. This final relation is the starting point for the study of azimuthal geodesics.

\section{Complete set of solutions}
\label{section:solutions}

Before studying azimuthal geodesics in detail, we briefly discuss the complete set of solutions of the geodesic equation in a FLRW spacetime. It is noteworthy that these can be integrated and the corresponding solutions only involve elementary functions.

\subsection{Angular motion}

Beginning with~\eqref{eq:geotheta} and~\eqref{eq:geophi}, one can eliminate the radial coordinate and the temporal coordinate to arrive at
\begin{equation}
    \dv{\theta}{\varphi} = 
    \sin{\theta}\sqrt{\frac{L^2}{L_z^2}\sin^2\theta-1}\,,
    \label{eq:diff1}
\end{equation}
which can be integrated to give
\begin{equation}
   \cot^2\negmedspace\theta = \left(\frac{L^2}{L_z^2}-1\right)
   \sin^2(\varphi-\varphi_0) 
   \label{eq:greatcircle}\,,
\end{equation}
where $\varphi_0$ is a constant of integration, fixed by the initial conditions. Equation~(\ref{eq:greatcircle}) describes arbitrary \emph{great circles}, and we note that for $L=L_z$ one finds $\theta=\pi/2$. For all $L \neq L_z$, the geodesic motion is three-dimensional with the angular momentum $L$ precessing around the $z$-axis. In fact, due to the spatial isotropy one can always choose $L=L_z$ and hence
the azimuthal geodesics are always great circles.

\subsection{Radial motion in polar angle form} 
By employing (\ref{eq:geor}) and (\ref{eq:geotheta}), we find
\begin{equation}
    \left(\dv{r}{\theta}\right)^2=-\frac{r^2 \left(k r^2-1\right) \left(L^2 \left(k r^2-1\right)+p^2
   r^2\right)}{L^2-L_z^2 \csc ^2(\theta )}\label{eq:rtheta}\,,
\end{equation}
which can be re-written as
\begin{equation}
    \left(\dv{r}{\theta}\right)^2=-r^2\left[ \left(k^2 L^2+k p^2\right)r^4+ \left(-2 k L^2-p^2\right)r^2+L^2\right]\frac{1}{L^2}\left(1-\frac{1} {\left(L/L_z\right)^2\sin^2(\theta)}\right)^{-1}\label{eq:rtheta2}.
\end{equation}
Now, \eqref{eq:rtheta2} can be solved by separation of variables, that is,
\begin{equation}
\arctanh\left(\sqrt{\frac{p^2}{L^2}\frac{r^2}{ \left(k r^2-1\right)}+1}\right)=\left[\arcsin\left(\frac{\cos{\theta}}{\sqrt{1-\left(L_z/L\right)^2}}\right)-\arcsin\left(\frac{\cos{\theta_0}}{\sqrt{1-\left(L_z/L\right)^2}}\right)\right], \label{eqn:intermediate}
\end{equation}
where $\theta_0$ is fixed by the initial conditions and depends, in fact, 
on the orbit's radius. If the orbit has a finite maximal radius (which is, for example, always true for $k=1$), it is convenient to choose $\theta_0=\pi-\arcsin(L/L_z)$ so that \eqref{eqn:intermediate} reads
\begin{equation}
    \arctanh\left(\sqrt{\frac{p^2}{L^2}\frac{r^2}{ \left(k r^2-1\right)}+1}\right)=\arcsin\left(\frac{\cos{\theta}}{\sqrt{1-\left(L_z/L\right)^2}}\right)+\frac{\pi}{2}\,.
\end{equation}
If the orbit's radius extends to infinity (which is possible, for example, when $k=0$ or $k=-1$), we choose $\theta_0=\pi/2$ such that \eqref{eqn:intermediate} becomes
\begin{equation}
   \arctanh\left(\sqrt{\frac{p^2}{L^2}\frac{r^2}{ \left(k r^2-1\right)}+1}\right)=\arcsin\left(\frac{\cos{\theta}}{\sqrt{1-\left(L_z/L\right)^2}}\right).
\end{equation}
\subsection{Radial motion in cosmological time} 
We can now consider \eqref{eq:geor} and \eqref{eq:geot} to find a relation for $r(t)$ or, equivalently, $t(r)$. In particular, we have
\begin{equation}
    \left(\dv{r}{t}\right)^2=-\frac{\left(k r^2-1\right) \left(L^2 \left(k r^2-1\right)+p^2 r^2\right)}{r^2
   a^2 \left(a^2 \mathcal{L} +k L^2+p^2\right)}\,.
\end{equation}
This equation is separable and we can find
\begin{multline}
    \int \frac{r \,\dd r}{\sqrt{-\left(k r^2-1\right) \left(L^2 \left(k r^2-1\right)+p^2 r^2\right)}}\\=\frac{1}{\sqrt{k(k L^2+p^2)}} \arctan(\sqrt{\frac{p^2}{\left(1-k r^2\right) \left(k L^2+p^2\right)}-1})+c_1
\end{multline}
To find the integral with respect to $t$, we need to specify $a(t)$. We consider two general forms for $a(t)$ which are usually discussed in this context.
First, consider a power law which arises e.g.\ for a matter or radiation dominated universe,
\begin{equation}
    a(t)=\alpha t^{\mu}\,,
\end{equation}
where $\alpha$ is an integration constant and $\mu \in \mathbb{R}^{+}$ is a dimensionless parameter. We therefore have
\begin{equation}
    \int \frac{\dd t}{a\sqrt{\left(a^2 \mathcal{L} +k L^2+p^2\right)}}=-\frac{t^{1-\mu } \sqrt{k L^2+p^2+\alpha ^2 \mathcal{L}  t^{2 \mu }} \,
   _2F_1\left(1,\frac{1}{2 \mu };\frac{\mu +1}{2 \mu };-\frac{t^{2 \mu } \alpha ^2
   \mathcal{L} }{k L^2+p^2}\right)}{\alpha  (\mu -1) \left(k L^2+p^2\right)}+c_2\,,
\end{equation}
and we note that for massless test particles ($\mathcal{L}=0$), we find the relation
\begin{equation}
    \frac{1}{\sqrt{k}} \arctan(\sqrt{\frac{p^2}{\left(1-k r^2\right) \left(k L^2+p^2\right)}-1})=\frac{t^{1-\mu }}{(\alpha -\alpha  \mu )}+c_3.
\end{equation}
Here $c_2$ and $c_3$ denote constants of integration. 

Lastly, we consider an exponential form, e.g.\ for a universe to be dominated by a cosmological constant  $\lambda > 0$, 
\begin{equation}
a(t)=\alpha \exp\left(\lambda t\right),
\end{equation}
where $\alpha$ denotes an integration constant. Then we have
\begin{equation}
    \int \frac{\dd t}{a\sqrt{\left(a^2 \mathcal{L} +k L^2+p^2\right)}}=-\frac{\mathrm{e}^{-\lambda t} \sqrt{k L^2+p^2+\alpha ^2 \mathcal{L}  \mathrm{e}^{2 \lambda  t}}}{\alpha 
   k \lambda  L^2+\alpha  \lambda  p^2}+c_2\,.
\end{equation}
We once again note that for $\mathcal{L}=0$, i.e.\, for a massless test particle, we obtain the relation
\begin{equation}
  \frac{1}{\sqrt{k}} \arctan(\sqrt{\frac{p^2}{\left(1-k r^2\right) \left(k L^2+p^2\right)}-1})= -\frac{\mathrm{e}^{-\lambda t}}{\alpha  \lambda}+c_3\,.
\end{equation}
In principle, this can be rewritten to find an explicit expression for the radial coordinate $r$ in terms of cosmological time $t$. Next, we will present a comprehensive study of the azimuthal geodesics.

\section{Azimuthal geodesics in closed FLRW spacetimes}\label{sec:azimuthalgeodesics}

We can use Eq.~\eqref{eq:friedmann1a} to derive the value of the
angle travelled by a particle on an azimuthal geodesic during one cycle of expansion and re-collapse for a closed FLRW universe in terms of $\beta_i$, for $i\in\{1,2\}$.

In what follows, we are concerned with azimuthal geodesics in the closed ($k=1$) FLRW spacetime, i.e. we are interested in motion along the azimuthal $\varphi$-direction only, i.e. at constant values of $r$ and $\theta$. Inspection of the components of the geodesic equation (\ref{eq:geo1})-(\ref{eq:geo4}) leads to the conclusion that we have to set $\theta=\pi/2$ and $r=1$. This implies that $L_z=L=a^2(t) \varphi'$.
By combining $\varphi'=\dot{\varphi}t'$ together with Eq.~(\ref{eq:azi1}), we find
\begin{align}
    \frac{{L_z}}{a^2} = 
    \dv{\varphi}{t}\left(\frac{L_z}{a}
    \sqrt{1-\frac{\mathcal{L}}{{L_z}^2}a^2}\right)\,.
    \label{azi5}
\end{align}
By separating the variables, we then obtain the expression for the angle travelled by a particle moving along an azimuthal geodesic during the expansion and subsequent re-collapse of
the universe. Assuming $a(0)=0$ and symmetry with respect to the value of the cosmic time $t=t_{\rm max}$ at which $a(t)$ attains its maximum value $a_{\rm max}$,  we find
\begin{align}
    \Delta\varphi = 2\int\limits_{0}^{t_{\rm max}}
    \frac{\dd t}{a\sqrt{1-\left(\mathcal{L}/L_z^2\right)a^2}}\,.
    \label{azi7}
\end{align}
By employing Eq.~(\ref{eq:friedmann1a}), this can be rewritten as
\begin{align}
    \Delta\varphi = 2\int\limits_{0}^{a_{\rm max}} 
    \left[a^2\left(1-\frac{\mathcal{L}}{L_z^2}a^2\right)\left(\beta_1 a^{-\gamma_1}+ \beta_2 a^{-\gamma_2}-1\right)\right]^{-1/2} \dd a \,,
    \label{azi8}
\end{align}
where the maximum value of the scale factor $a_{\rm max}$ satisfies the conditions $\dot{a}=0$ in Eq.~(\ref{eq:friedmanna}) and $\ddot{a} < 0$ in Eq.~(\ref{eq:friedmannb}).

\subsection{Massless particles in two-fluid cosmologies}\label{sec:masslesstwofluid}
As we can see from \eqref{eq:rhoanda}, the energy-momentum density evolves differently according to the equation of state parameter. In the case of a universe filled with two barotropic fluids, this means that one of the fluids may prevail for some time in the universe's evolution. This approach allows for a more comprehensive understanding of the dynamics of the universe. For a review on multifluid cosmologies see e.g.~\cite{Faraoni:2021opj}.
Azimuthal geodesics for light-like particles in single-fluid cosmologies have been discussed previously \cite{Boehmer:2024kdt}. Here, we return to the case of massless particles and now extend the approach to the two-fluid case. 

For convenience, we use the rescaling
\begin{equation}
a \mapsto \left(\beta_1\right)^{1/\gamma_1} a \,, \quad  \text{and}\quad 
    t \mapsto \left(\beta_1\right)^{1/\gamma_1} t\,,
\end{equation}
such that Eq.~(\ref{eq:friedmann1a}) can be re-written in terms of $\xi:=\beta_2/(\beta_1)^{\gamma_2/\gamma_1}$, and reads
\begin{equation}
    \left(\dot{a}\right)^2= a^{-\gamma_1}+\xi a^{-\gamma_2}-1\,. \label{eq:friedmann1}  
\end{equation}
Moreover, since $\mathcal{L}=0$ is for null geodesics, Eq.~(\ref{azi8}) becomes
\begin{align}
    \Delta\varphi = 2\int\limits_{0}^{a_{\rm max}}
   \left[a^2\left(a^{-\gamma_1}+ \xi a^{-\gamma_2}-1\right)\right]^{-1/2}\dd a \, .
    \label{azi9}
\end{align}
This integral can be given in terms of elementary functions only for specific choices of $\gamma_1$ and $\gamma_2$. This is related to the Theorem of Chebyshev, see Appendix \ref{sub:C_theorem}. 

\subsubsection{Dust and radiation}
\label{subsection:dust_radiation}
For a universe filled with dust ($w_1=0$, i.e.\ $\gamma_1=1$) and 
radiation ($w_2=1/3$, i.e.\ $\gamma_2=2$), Eq.~(\ref{azi9}) yields
\begin{align}
\label{eq:delta_phi_dust_radiation}
    \Delta \varphi=2\int_{0}^{a_{\rm max}}\frac{\dd a}{\sqrt{-a^2+a+\xi}}
    =2\left[\arcsin(\frac{a-\frac{1}{2}}{\sqrt{\frac{1}{4}+\xi}})\right]_{0}^{a_{\rm max}}.
\end{align}
The maximum value of the scale factor can be found from Eq.~(\ref{eq:friedmann1}) and is given by
\begin{equation}
    a_{\rm max}=\frac{1}{2} +\sqrt{\frac{1}{4}+ \xi} \,.
\end{equation}
Inserting this into Eq.~(\ref{eq:delta_phi_dust_radiation}) gives
\begin{equation}
    \Delta\varphi =\pi +2\arcsin\left(\frac{1}{\sqrt{4\xi+1}}\right) \ .
\end{equation}
In Fig.\ \ref{fig:twofluid_dustradi}, we show $\Delta\varphi$ as a function of $\xi=\beta_2/\beta_1^2$. The limit $\xi\to 0$ corresponds to a universe that contains only dust, for which we obtain $\Delta\varphi=2\pi$, as expected. In the limit $\xi\rightarrow \infty$, which corresponds to a universe that contains only radiation, we find  the known result $\Delta\varphi \rightarrow \pi$. 


\subsubsection{Radiation and stiff matter}
\label{subsection:stiff_radiation}

For a universe filled with radiation ($w_1=1/3$, i.e.\ $\gamma_1=2$) and 
stiff matter ($w_2=1$, i.e.\ $\gamma_2=4$), we find from Eq.~(\ref{azi9}) 
\begin{align}
\label{eq:delta_phi_stiff_radiation}
    \Delta\varphi = 2\int\limits_{0}^{a_{\rm max}}
    \frac{\dd a}{\sqrt{1+ \xi a^{-2}-a^2}} = \left[\arcsin(\frac{a^2 -\frac{1}{2}}{\sqrt{\frac{1}{4}+\xi}})\right]_{0}^{a_{\rm max}} \ .
\end{align}
The maximum value of the scale factor can be found from (\ref{eq:friedmann1}) and reads
\begin{equation}
    a_{\rm max}=\sqrt{\frac{1}{2} +\sqrt{\frac{1}{4}+ \xi}} \ .
\end{equation}
Inserting this into Eq.~(\ref{eq:delta_phi_stiff_radiation}) gives
\begin{equation}
    \Delta\varphi =\frac{\pi}{2} +\arcsin\left(\frac{1}{\sqrt{4\xi+1}}\right) \ .
\end{equation}

In Fig.~\ref{fig:twofluid_radistiff}, we show $\Delta\varphi$ in function of $\xi=\beta_2/\beta_1^2$. The limit $\xi\to 0$ corresponds to a universe that contains only radiation, for which $\Delta\varphi=\pi$. In the limit $\xi\rightarrow \infty$, i.e.\ the limit
corresponding to a universe filled only with stiff matter, we obtain $\Delta\varphi\to\pi/2$. 

Note that, mathematically, the case discussed here is similar to that for a dust and radiation filled universe, see Sec.~\ref{subsection:dust_radiation}. The reason is that the ratio $\gamma_2/\gamma_1=2$ in both cases. 

\subsubsection{Dust and stiff matter}

Here, we consider a universe filled with dust ($w_1=0$, i.e.\ $\gamma_1=1$) and stiff matter ($w_2=1$, i.e.\ $\gamma_2=4$). Unlike the cases studied in
Sec.~\ref{subsection:dust_radiation} and Sec.~\ref{subsection:stiff_radiation}, respectively, we now have $\gamma_2/\gamma_1=4$ and $\xi=\beta_2/\beta_1^4$.
The expression (\ref{azi9}) then becomes
\begin{equation}
    \Delta\varphi=2\int\limits_0^{a_{\rm max}}\frac{a}{\sqrt{a^3+\xi-a^4}}\dd a\,. \label{eqn:azdistduststiff}
\end{equation}
and $a_{\rm max}$ is one of the roots of the polynomial
\begin{equation}
a^4 - a^3 - \xi=0 \,.
\end{equation}
This is an integral which cannot be given in terms of elementary functions, but is elliptic.

\begin{figure}[p]
	\centering
	\begin{subfigure}{0.55\textwidth}
		\includegraphics[width=\textwidth]{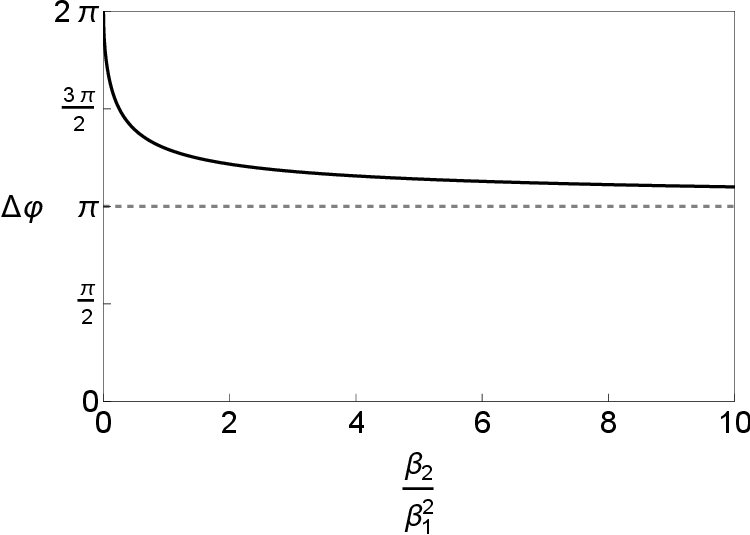}
        \caption{Dust and radiation.}
        \label{fig:twofluid_dustradi}
	\end{subfigure}\\
	\begin{subfigure}{0.55\textwidth}
		\includegraphics[width=\textwidth]{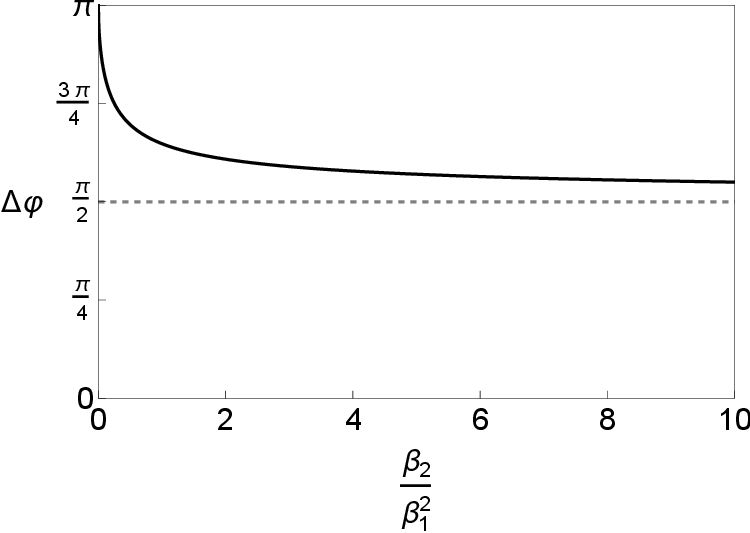}
		\caption{Radiation and stiff matter.}
        \label{fig:twofluid_radistiff}
	\end{subfigure}\\
	\begin{subfigure}{0.55\textwidth}
	    \includegraphics[width=\textwidth]{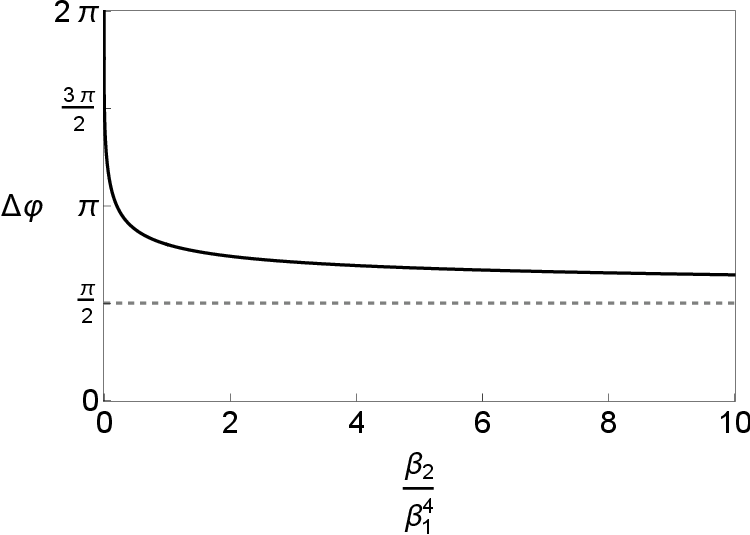}
        \caption{Dust and stiff matter.}
        \label{fig:twofluid_duststiff}
         \end{subfigure}
	\caption{We show $\Delta\varphi$ in function of  $\xi=\beta_2/\beta_1^{\gamma_2/\gamma_1}$ for a massless particle moving in a universe filled with two fluids.}
	\label{fig:subfigures}
\end{figure}

The analytic expression for $\Delta\varphi$ in this case is quite complicated since it involves a combination of the incomplete elliptic integrals of the first kind $F[\phi,m]$ and of the third kind $\Pi[n,\phi,m]$ (see Appendix \ref{subsec:elliptic} for more details), containing all the roots of $a^4-a^3-\xi=0$. We remark that the discriminant of this polynomial is given by $\Delta=-\xi ^2 (256 \xi +27)<0$, since $\xi>0$, this means that the polynomial has two real roots (one positive and one negative), and two complex conjugate roots. We denote these roots by $a_i$, for $i\in\left\{1,2,3,4\right\}$, and $a_{\rm max}$ is the largest positive real root, and find
\begin{equation}
\Delta\varphi= \left[\frac{2}{\sqrt{\left(a_1-a_3\right)\left(a_2-a_4\right)}}
\left(a_2\, F\left[\phi, m\right]+\left(a_1-a_2\right)\Pi\left[n, \phi, m\right] 
\right) \right]_0^{a_{\rm max}},
\end{equation}
where
\begin{equation}
\phi=\arcsin\left[\sqrt{\frac{\left(a-a_1\right) \left(a_2-a_4\right)}{\left(a-a_2\right)
\left(a_1-a_4\right)}}\right], \quad 
n= \frac{a_1-a_4}{a_2-a_4} \,, \quad
m=\frac{\left(a_2-a_3\right)
\left(a_1-a_4\right)}{\left(a_1-a_3\right)
\left(a_2-a_4\right)}\,.
\end{equation}
Noting that, for $a_{\rm max}=a_4$, the upper integration boundary gives the complete elliptic integrals of the first and third kind. We provide a plot of $\Delta\varphi$ as a function of $\xi$ in Fig.~\ref{fig:twofluid_duststiff}. The behaviour of the function $\Delta\varphi$ for small and large values of $\xi$ can be easily examined. 
As $\xi\rightarrow 0$, we have
\begin{equation}
    \Delta\varphi =2\int\limits_0^{t_{\rm max}}\frac{\dd t}{a(t)}=2\int\limits_0^{a_{\rm max}}\frac{\dd a}{\sqrt{a-a^2}}\,
\end{equation}
where we note that $a_{\rm max}\rightarrow 1$ in this limit. We thus find that for $\xi\rightarrow 0$, 
\begin{equation*}
    \Delta\varphi =2\int_0^{1} \frac{\dd a}{\sqrt{a-a^2}}=2\pi\,,
\end{equation*}
which is indeed consistent with the fact that as $\xi \rightarrow 0$, we are in the limit $\beta_2\rightarrow 0$.

For large values of $\xi$, it is more helpful to return to the integral form containing $\beta_1$ and $\beta_2$, and note that as $\beta_1\rightarrow 0$, the integral becomes
\begin{equation*}
    \Delta \varphi = 2 \int_0^{\beta_2^{1/4}} \frac{a}{\sqrt{\beta_2 -a^4}}\dd a = \frac{\pi}{2}.
\end{equation*}

\subsection{Masless particle in a fluid with a cosmological constant}\label{sec:cosmologicalconst}
We will now discuss the azimuthal geodesics of a massless particle in a universe filled by matter, radiation, or stiff matter, together with a cosmological constant. To this end, we note that, for a cosmological constant $\Lambda$, we can use \eqref{eq:friedmann1a} where we set $\beta_2\equiv \Lambda/3$ and $\gamma_2=-2$, or, equivalently, $w_2=-1$, effectively treating the cosmological constant as a fluid. In the following, we also set $\beta_1=\beta$, $\gamma_1=\gamma$, and, thus, $\xi=\Lambda\beta^{2/\gamma}/3$. Note that $\xi \gtrless 0$ for $\Lambda \gtrless  0$. 

Before proceeding with our analysis, we seek the condition that the solution $a_{\rm max}$ has to satisfy in order for it to be a positive real maximum. Equations~\eqref{eq:friedmann1a} and~\eqref{eq:friedmannb} read
\begin{align}
    \dot{a}^2&=a^{-\gamma}+ \xi a^2-1\, , \label{eq:l3}\\
   a\ddot{a}&=-\frac{1}{2}\gamma a^{-\gamma}+ \xi a^2=-\frac{1}{2}\gamma a^{-\gamma}+ \xi a^2\,. \label{eq:l5}
\end{align}

\paragraph{Positive cosmological constant.}

First we note that if $a$ is a (positive) maximum, then from Eq.~\eqref{eq:l3}, we can write
\begin{equation}
    a^{-\gamma}=1-\xi a^2\,, \label{eq:l6}
\end{equation}
whence Eq.~\eqref{eq:l5} reads
\begin{equation}
     a\ddot{a}=-\frac{1}{2}\gamma\left(1-\xi a^2\right)+\xi a^2=-\frac{\gamma}{2}+\left(\frac{\gamma}{2}+1\right)\xi a^2. 
\end{equation}
To ensure the existence of a (positive) maximum we require $\ddot{a}<0$, which implies
\begin{equation}
    -\frac{\gamma}{2}+\left(\frac{\gamma}{2}+1\right)\xi a^2<0 \iff 0<a<\frac{1}{\sqrt{\xi}}\sqrt{\frac{\gamma}{\gamma+2}}\equiv \frac{1}{\sqrt{\xi}}\sqrt{\frac{1+3w}{3(1+w)}}. \label{eq:l8}
\end{equation}
In order to be able to make some more concrete statements, we need to fix the value of $w$, and thus $\gamma$, and discuss the number of roots of Eq.~\eqref{eq:l6}.

\paragraph{Negative cosmological constant} We note that Eq.~\eqref{eq:l5} does not yield a constraint for a maximum as $a\ddot{a}<0$ for all $a$, $\gamma$, and $\xi$. However, this does not constitute a problem as it will be straightforward to identify the maximum in the specific scenarios.
\subsubsection{Dust and cosmological constant}
\label{sec:CCmatterdom}
We now discuss the case of a universe filled with dust ($w_1=0$, i.e.\ $\gamma_1=1$)
and a cosmological constant ($w_2=-1$, i.e.\ $\gamma_2=-2$). Equation (\ref{azi9}) then reads
\begin{equation}
    \Delta \varphi=2\int\limits_{0}^{a_{\rm max}}\frac{\dd a}{\sqrt{a+\xi a^{4}-a^2}}\,,\label{eq:l13} 
\end{equation}
where $\xi=\Lambda \beta^2/3$.
By letting $a=1/(x+1/3)$, which implies $\dd a=-1/(x+1/3)^2$, Eq.~\eqref{eq:l13} becomes
\begin{equation}
    \Delta\varphi = -2\int\limits_{\infty}^{x_{\rm max}} \frac{\dd x}{\sqrt{x^3-x/3
    +\xi-2/27}}=4\int\limits_{x_{\rm max}}^{\infty}\frac{\dd x}{\sqrt{4x^3-4x/3+ 4\xi-8/27}}\,,
\end{equation}
where $x_{\rm max}=a_{\rm max}^{-1}-1/3$. We have thus obtained an elliptic integral in Weierstrass form (see Appendix \ref{subsec:weierstrassP} for more details). Hence,
\begin{equation}
    \Delta\varphi=4\wp^{-1}\left(\frac{1}{a_{\rm max}}-\frac{1}{3};\frac{4}{3};\frac{8}{27}- 4\xi\right), \label{eq:l15}
\end{equation}
where $\wp^{-1}$ is the inverse Weierstrass $\wp$-function. 
The value of $a_{\rm max}$ can be found from Eq.~(\ref{eq:friedmann1}) by determining the roots of the third-degree polynomial in the scale factor $a$, that is,
\begin{equation}
    \xi a^3-a+1=0\,.\label{eq:l10}
\end{equation}
In the following, we will discuss the cases of the positive and the negative cosmological constant separately. The discriminant of the polynomial on the left-hand side of Eq.~\eqref{eq:l10} is given by $\Delta=(4-27\xi)\xi$.

\paragraph{Positive cosmological constant ($\xi > 0$)}
One can easily see that the depressed cubic polynomial in Eq.~\eqref{eq:l10} has at least one negative real root and two complex conjugate roots if $\Delta<0$, i.e.\ $\xi>4/27$; and it has three real roots if $\Delta\geq0$, i.e.\ $0<\xi<4/27$. We can therefore restrict our attention to the case $0<\xi<4/27$, as other values would not lead to a closed universe. In the present case, from Eq.~\eqref{eq:l8}, we have an additional constraint on the value of $a_{\rm max}$ 
\begin{equation}
    0<a_{\rm max}<1/\sqrt{3\xi}\ . 
\end{equation}
It can than be verified that the value of $a_{\rm max}$ is given by
\begin{equation}
    a_{\rm max}=\frac{\sqrt[3]{-1} \left(2 \sqrt[3]{-3} \xi -\sqrt[3]{2} \left(\xi ^{3/2} \sqrt{81 \xi
   -12}-9 \xi ^2\right)^{2/3}\right)}{6^{2/3} \xi  \sqrt[3]{\xi ^{3/2} \sqrt{81 \xi
   -12}-9 \xi ^2}}\,.
   \label{eq:amaxdustcos}
\end{equation}
Note that although, at first glance, Eq.~\eqref{eq:amaxdustcos} appears to be complex (due to the square roots of negative terms), it is indeed real. In Fig.~\ref{fig:cc_matterdom}, we show the value of $\Delta\varphi$ as a function of $\xi$. 
We note that as $\xi\rightarrow 0$, $\Delta\varphi\rightarrow 2\pi$, while for $\xi\rightarrow4/27$, $\Delta\varphi\rightarrow\infty$. This can also be verified by setting $\xi=4/27$ in the initial integral, Eq.~\eqref{eq:l13}. Physically, this means that, if the positive cosmological constant is dominant in the universe, closed universe solutions are no longer possible and, hence, the angular distance travelled by a massless particle on an azimuthal geodesic diverges. 

\paragraph{Negative cosmological constant ($\xi < 0$)}
 
    We note that the discriminant, $\Delta$, is always negative in this case.
    This means that there is only one positive real root, which is our maximum, and two complex conjugate roots. In other words, for a negative cosmological constant, the universe always exhibits oscillatory behaviour. The value of $a_{\rm max}$ is thus given by
\begin{equation}
    a_{\rm max}=\frac{\sqrt[3]{2} \left(\sqrt{3} \sqrt{\xi ^3 (27 \xi -4)}-9 \xi ^2\right)^{2/3}+2
   \sqrt[3]{3} \xi }{6^{2/3} \xi  \sqrt[3]{\sqrt{3} \sqrt{\xi ^3 (27 \xi -4)}-9 \xi ^2}}\,.
\end{equation}
We remark that, as $\xi\to -\infty$, $\Delta\varphi\rightarrow 0$.

\subsubsection{Radiation and cosmological constant}
We now turn our attention to the case of a universe filled with radiation ($w_1=1/3$, i.e.\ $\gamma_1=2$) and a cosmological constant ($w_2=-1$, i.e.\ $\gamma_2=-2$). This gives
\begin{equation}
    \Delta \varphi=2\int\limits_{0}^{a_{\rm max}}\frac{\dd a}{\sqrt{1+\xi a^{4}-a^2}}\,,\label{eq:l19}
\end{equation}
where $\xi=\Lambda \beta/3$,
and $a_{\rm max}$ is computed from 
\begin{equation}
    \xi a^4-a^2+1=0. \label{eq:l17}
\end{equation}
Equation (\ref{eq:l17}) is bi-quadratic in $a$, and the maximum real root is given by
\begin{equation}
\label{eq:root1}
a_{\rm max}=\frac{\sqrt{2}}{\sqrt{\sqrt{1-4 \xi }+1}}\,,
\end{equation}
for all values of $\xi$. The value of $\Delta\varphi$ then is 
\begin{equation}
    \Delta\varphi =\frac{2\sqrt{2}}{\sqrt{1+\sqrt{1-4 \xi }}}K\left(\frac{1-\sqrt{1-4 \xi }}{1+\sqrt{1-4 \xi
   }}\right), \label{eq:l20}
\end{equation}
where $K$ is the complete elliptic integral of the first kind (see Appendix \ref{subsec:elliptic}).
\paragraph{Positive cosmological constant ($\xi > 0$)}
It is clear that we require 
\begin{equation}
    1-4\xi>0 \iff 0<\xi<1/4.
\end{equation}
Note that the root (\ref{eq:root1}) satisfies the condition $0<a<1/\sqrt{2\xi}$, from Eq.~\eqref{eq:l8}, and hence is, indeed, a maximum.
The dependence of $\Delta\varphi$ on $\xi$ is shown in 
Fig.~\ref{fig:cc_raddom}.  In the limit $\xi\rightarrow 0$, we see that 
$\Delta\varphi=\pi$ as expected. As $\xi\rightarrow1/4$, $\Delta\varphi\rightarrow\infty$. The divergence again relates to the fact that,
when the positive cosmological constant becomes too large, closed universe solutions no longer exist. 

\paragraph{Negative cosmological constant ($\xi < 0$)}
In this case, we have again no restriction on $\xi$, and the universe always exhibits oscillatory behaviour. As $\xi\to-\infty$, then $\Delta\varphi$ decays to zero.

\subsubsection{Stiff matter and cosmological constant\label{sec:poscosmconststiffmatter}}
For stiff matter ($w_1=1$, i.e.\ $\gamma_1=4$) and a cosmological constant ($w_2=-1$, i.e.\ $\gamma_2=-2$), we get
\begin{equation}
    \Delta\varphi=2\int\limits_0^{a_{\rm max}} \frac{a}{\sqrt{1+\xi a^6-a^4}}\dd a\,, \label{eq:sm2}
\end{equation}
where $\xi=\Lambda \sqrt{\beta}/3$. The maximum value of $a$ corresponds to one of the roots of 
\begin{equation}
    \xi a^6-a^4 + 1 = 0\,,\label{eq:sm1}
\end{equation}
and we denote it by $a_{\rm max}$, as usual. 
Let us perform the substitution
\begin{equation}
    y=3 \xi a^2- 1 \iff \dd y=6 \xi a \dd a \,,
\end{equation}
which allows us to re-write Eq.~\eqref{eq:sm2} as
\begin{align}
        \Delta\varphi &= \operatorname{sgn}(\xi)2\sqrt{3} \int\limits_{-1}^{ 3 \xi a_{\rm max}^2-1} \frac{ \dd y}{\sqrt{4y^3-12y+4(27\xi^2-2)}}\\
        &=\operatorname{sgn}(\xi)2\sqrt{3}\left[\wp^{-1}\left(y;12;-4\left(27\xi^2-2\right)\right)\right]_{-1}^{3 \xi a_{\rm max}^2-1}\,.
\end{align}
Note that if we take the limit $\xi\rightarrow 0$ in Eq.~\eqref{eq:sm2}, we can easily find the value of $\Delta\varphi$ which corresponds to that of a universe without a cosmological constant. Indeed, from Eq.~\eqref{eq:sm1} as $\xi\to 0$, the maximum value of $a_{\rm max}=1$, and we have
\begin{equation*}
    \Delta\varphi=\int\limits_0^{a_{\rm max}} \frac{2a }{\sqrt{1-a^4}}\dd a=\frac{\pi}{2}\,.
\end{equation*}

\paragraph{Positive cosmological constant ($\xi > 0$)} 
We need to find a condition on $\xi$ which guarantees that the polynomial in Eq.~\eqref{eq:sm1} has a positive real root. This polynomial can be reduced to a cubic polynomial by letting $x=a^2$, which is $\xi x^3-x^2+1=0$. A positive real root of Eq.~\eqref{eq:sm1} can arise only as the positive square root of a positive real root $x$ of $\xi x^3-x^2+1=0$. It is easy to check that it has exactly one negative real root, and so it has a positive real root if and only if it has more than one real root, i.e.\ if and only if it has non-negative discriminant. Its discriminant is $\Delta=4-27\xi^2$, which is non-negative only if  $\xi\le2/\left(3\sqrt{3}\right)$. 

\begin{figure}
	\centering
	\begin{subfigure}{0.54\textwidth}
		\includegraphics[width=\textwidth]{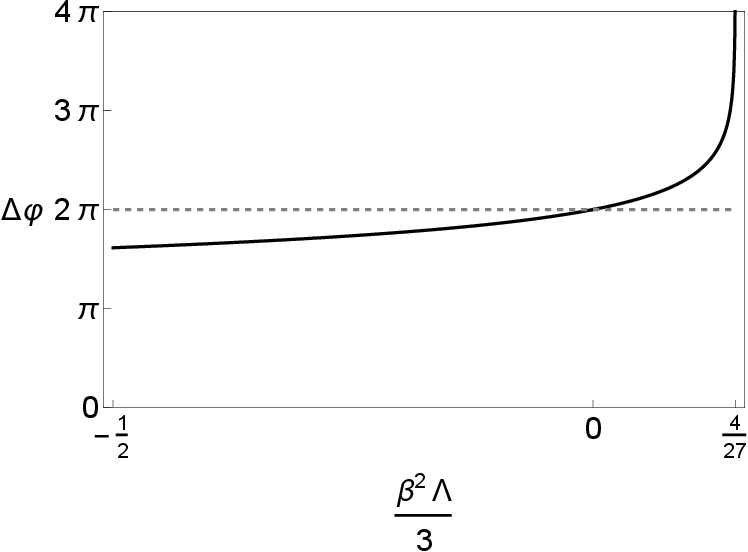}
        \caption{Dust.}
        \label{fig:cc_matterdom}
	\end{subfigure}\\
	\begin{subfigure}{0.54\textwidth}
		\includegraphics[width=\textwidth]{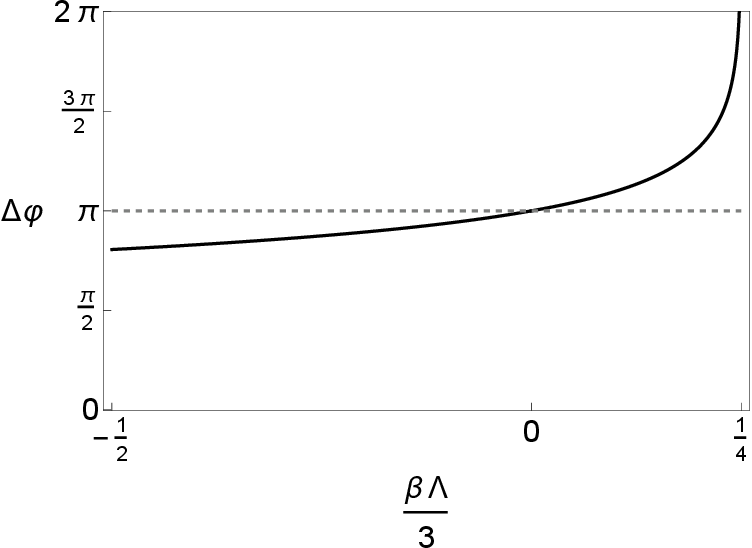}
		\caption{Radiation.}
        \label{fig:cc_raddom}
	\end{subfigure}\\
	\begin{subfigure}{0.54\textwidth}
	    \includegraphics[width=\textwidth]{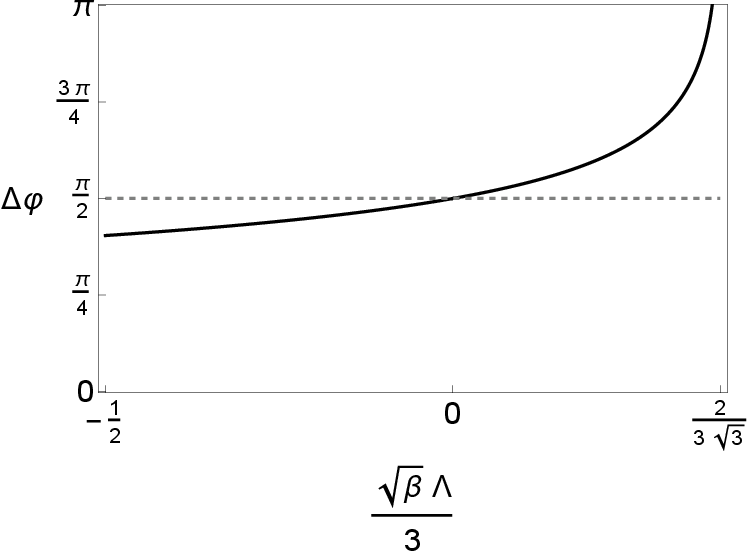}
        \caption{Stiff matter.}
        \label{fig:cc_stiffdom}
         \end{subfigure}
	\caption{We show the angular distance $\Delta\varphi$ as a function of $\xi=\Lambda\beta^{2/\gamma}/3$ for a universe filled with cosmological constant and a second perfect, barotropic fluid.}
	\label{fig:subfigures2}
\end{figure}

Therefore, when $0<\xi\le2/\left(3\sqrt{3}\right)$, the polynomial in Eq.~\eqref{eq:sm1} has two positive roots and two negative roots (these will be repeated roots when $\xi=2/\left(3\sqrt{3}\right)$). Further, we have the constraint from Eq.~\eqref{eq:l8}.
It can be verified that the value of $a_{\rm max}$ is given by
\begin{equation}
   a_{\rm max}=\frac{\sqrt{2^{2/3} \left(-1-\mathrm{i} \sqrt{3}\right) \delta ^{2/3}+4 \sqrt[3]{\delta }+2 \mathrm{i}
   \sqrt[3]{2} \left(\sqrt{3}+\mathrm{i}\right)}}{2 \sqrt{3} \sqrt[6]{\delta } \sqrt{\xi }}\,, \label{eq:amaxstiffmatter}
\end{equation}
where $\delta=-27 \xi ^2+3 \xi \sqrt{3} \sqrt{27 \xi ^2-4} +2$. As in the case of Eq.~(\ref{eq:amaxdustcos}), this is real-valued, despite appearing to be complex. Once again, this is a constraint on the positive cosmological constant. For larger values of $\xi$, and hence of the cosmological constant, we do not have re-collapsing universes. In Fig.~\ref{fig:cc_stiffdom}, we show $\Delta\varphi$ as a function of $\xi$. 

\paragraph{Negative cosmological constant ($\xi < 0$)}
Again, there is no restriction on $\xi$ and the only real positive root is given by
\begin{equation}
    a_{\rm max}=\frac{\sqrt{(-2)^{2/3} \delta ^{2/3}+2 \sqrt[3]{\delta }-2 \sqrt[3]{-2}}}{\sqrt{6}
   \sqrt[6]{\delta } \sqrt{\xi }}\,,
\end{equation}
where $\delta$ is defined as above.
As $\xi\to-\infty$, then $\Delta\varphi$ decays to zero.

\subsection{Massive particles in single-fluid cosmologies}\label{sec:massiveone}

We shift our focus to the case of a massive particle in a closed FLRW universe, i.e.\ we choose $\mathcal{L}=-1$ in (\ref{azi7}) and (\ref{azi8}). We restrict our attention to single-fluid cosmologies, i.e.\ we set $\beta_1=\beta$, $w_1=w$, and $\gamma_1=\gamma$, while $\beta_2=w_2=\gamma_2=0$. In this context, we do not study the case of a cosmological constant, since the physically relevant case of a positive cosmological constant does not lead to closed universe solutions. Recall that for dust, radiation, and stiff matter, respectively, the maximum value of the scale factor $a(t)$ is given by 
\begin{equation}
    \label{eqn:amax1}
    a_{\mathrm{max}}=\left(\beta\right)^{1/\gamma}\,,
\end{equation}
as shown in \cite{Boehmer:2024kdt}.

\subsubsection{Dust}
Here $w=0$, i.e.\ $\gamma=1$, and consequently the integral to be evaluated reads
\begin{equation}
    \Delta \varphi =2\int_0^{\beta} \frac{\dd a}{a\sqrt{\beta a^{-1}-1}\sqrt{1+L_z^{-2}a^2}}=2\int_0^{\beta}\frac{\dd a}{\sqrt{a\left(\beta-a\right)\left(1+L_z^{-2}a^2\right)}}.
\end{equation}
Recall that, in Sec.~\ref{sec:azimuthalgeodesics}, we assumed $L_z>0$ to derive an expression for $\Delta \varphi$, and $\beta>0$ as it is an energy density. Note also that the integral is elliptic. We hence obtain
\begin{equation}
    \Delta\varphi=4\sqrt{\frac{L_z}{\sqrt{L_z^2+\beta
   ^2}}}K\left(\frac{1}{2}-\frac{1}{2}\frac{L_z}{ \sqrt{L_z^2+\beta
   ^2}}\right), \label{eq:matterdommassive}
\end{equation}
where $K$ is the complete elliptic integral of the first kind. 
In Fig.~\ref{fig:massivematterdom}, we show $\Delta\varphi$ as a function of the ratio between angular momentum and energy density $L_z/\beta$.
We find that as $L_z\rightarrow\infty$ then $\Delta\varphi \rightarrow 2\pi$. This is indeed consistent with the well-known result and the more general one we obtained in our recent work.

\subsubsection{Radiation}
In this case $w=1/3$, i.e.\ $\gamma=2$, and hence
\begin{equation}
    \Delta\varphi=2\int_0^{\sqrt{\beta}} \frac{\dd a}{a\sqrt{\beta a^{-2}-1}\sqrt{1+L_z^{-2}a^2}}=2\int_0^{\sqrt{\beta}}\frac{\dd a}{\sqrt{\left(\beta-a^2\right)\left(1+L_z^{-2}a^2\right)}}=2 K\left(-\frac{\beta }{L_z^2}\right)\,,
\end{equation}
provided that $\beta>0$. This integral is again elliptic.
In Fig.~\ref{fig:massiveradiationdom} we show $\Delta\varphi$ as a function of
$L_z/\sqrt{\beta}$. For $L_z\rightarrow\infty$ we find $\Delta\varphi\rightarrow \pi$, as expected.

\subsubsection{Stiff matter}
\begin{figure}[p]
	\centering
	\begin{subfigure}{0.55\textwidth}
		\includegraphics[width=\textwidth]{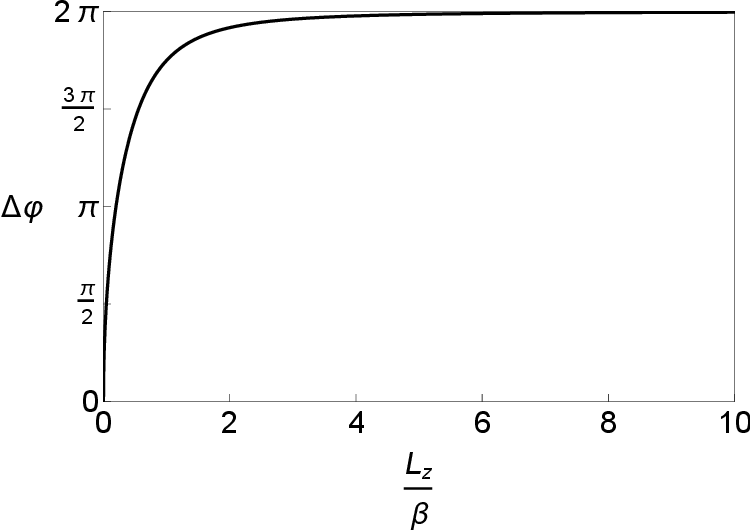}
        \caption{Dust.}
        \label{fig:massivematterdom}
	\end{subfigure}\\
	\begin{subfigure}{0.55\textwidth}
		\includegraphics[width=\textwidth]{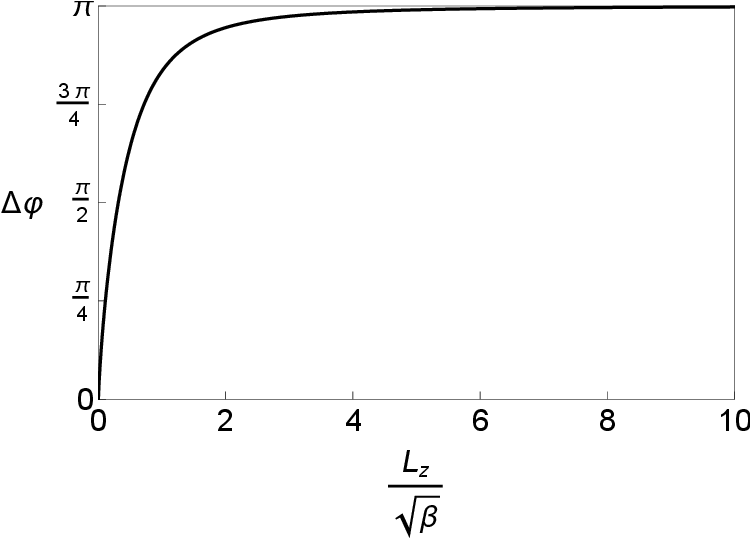}
		\caption{Radiation.}
        \label{fig:massiveradiationdom}
	\end{subfigure}\\
	\begin{subfigure}{0.55\textwidth}
	    \includegraphics[width=\textwidth]{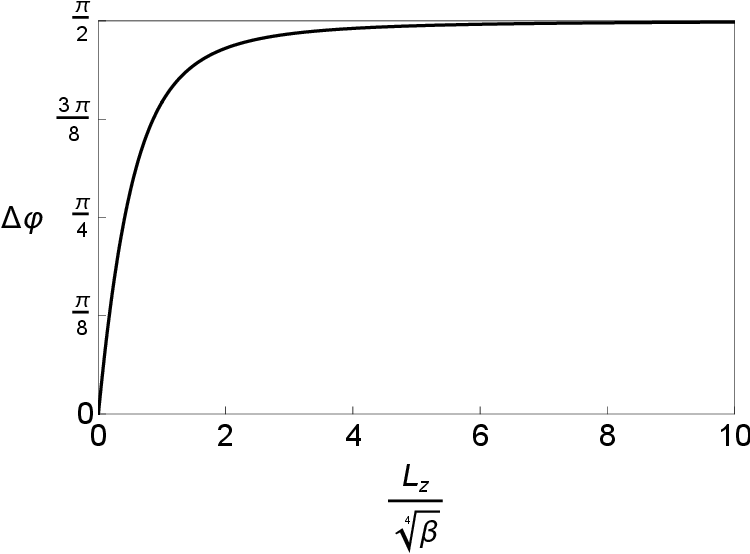}
        \caption{Stiff matter.}
        \label{fig:massivestiffdom}
         \end{subfigure}
	\caption{We show the angular distance $\Delta\varphi$ travelled by a massive test particle in a closed universe filled with a single perfect barotropic fluid. Here $L_z$ is the angular momentum of the test particle.}
	\label{fig:subfigures3}
\end{figure}

For $w=1$, i.e.\ $\gamma=4$, we have
\begin{align}
     \Delta\varphi&=2\int_0^{\sqrt[4]{\beta}} \frac{\dd a}{a\sqrt{\beta a^{-4}-1}\sqrt{1+L_z^{-2}a^2}}=2\int_0^{\beta^{1/4}}\frac{a}{\sqrt{\left(\beta -a^4\right)\left(1+L_z^{-2}a^2\right)}}\dd a\\
     &=
    \frac{L_z}{\sqrt{\sqrt{\beta
   }+L_z^2}}K\left(\frac{2 \sqrt{\beta }}{L_z^2+\sqrt{\beta }}\right)-\frac{1}{2}\frac{\sqrt{\beta }}{L_z^2}\,
   _3F_2\left(\frac{3}{4},1,\frac{5}{4};\frac{3}{2},\frac{3}{2};\frac{\beta }{L_z^4}\right),
\end{align}
provided $\beta > 0$. Here $_3F_2$ is the generalised hypergeometric function (see \ref{subsec:hypergeom} for more details).
In Fig.~\ref{fig:massivestiffdom}, we show $\Delta\varphi$ as a function of
$L_z/\sqrt[4]{\beta}$. In the limit $L_z\rightarrow \infty$, this returns $\Delta\varphi \rightarrow \pi/2$, which is consistent with our previous results. 

\section{Discussion and Conclusions}\label{sec:discandconc}

The study of geodesics is central to cosmology, as all electromagnetic signals travel along them. Consequently, they are important in the definition of cosmological distances and yield, for example, the distance-redshift relations. Moreover,
for high frequencies and in the so-called geometrical optics' limit, gravitational wave propagation can be studied using null geodesics \cite{Misner:1973prb}. In \cite{Fier:2021fbt}, this was recently used to describe gravitational waves produced by some astrophysical sources (in the primordial universe) and propagating through a non-empty universe. In this paper, we have presented the full set of solutions to the geodesic equation in FLRW spacetimes. As an application, we have extended our previous work~\cite{Boehmer:2024kdt} to find a closed formula for the azimuthal distance travelled by a massless or massive test particle, respectively, in a closed universe during one cycle of expansion and re-collapse. While an FLRW model does not allow closed universe solutions for a universe solely filled with a positive cosmological constant $\Lambda > 0$, two-fluid cosmologies allow us to include $\Lambda$. This is, of course, possible as long as the positive cosmological constant does not dominate over the second fluid in such a way as to render closed universe solutions impossible. In our computations, we find that, as expected, $\Delta \varphi$ diverges when we reach this limit. 
When considering a massless test particle moving in two-fluid cosmologies, we find that the integrals determining $\Delta\varphi$ can be solved in terms of elementary functions only for a universe filled with (a) dust and radiation or (b) radiation and stiff matter (see Appendix~{\ref{sub:C_theorem}}).

All other cases lead to elliptic integrals. Extending these results to a massive test particle in a one-fluid model, we find that for dust and radiation the value of $\Delta\varphi$ is given in terms of the complete elliptic integral of the first kind $K(m)$ with elliptic modulus $m$ depending on the energy density of the fluid and the angular momentum of the particle. For stiff matter, $\Delta\varphi$ is a sum of such an elliptic integral and a generalised hypergeometric function $_3F_2$ with argument given in terms of the energy density of the stiff matter and
the angular momentum of the test particle.

In principle, this work could be extended to other possible set-ups, e.g.~to the motion of a massive test particle in two-fluid cosmologies. As suggested by the general expression for $\Delta\varphi$
given here, this would likely involve hyperelliptic integrals or would require numerical tools. Moreover, it would be interesting to see whether the geometrical optics' limit provides a good approximation to the motion of high frequency gravitational waves through the universe using our solutions for the null geodesics.

\section*{Acknowledgments}
Antonio d'Alfonso del Sordo is supported by the Engineering and Physical Sciences Research Council EP/R513143/1 \& EP/T517793/1.

\appendix

\section{Special functions}

Special functions arise as indefinite integrals which cannot be expressed in terms of elementary functions and are ubiquitous in mathematical physics. For an overview of many special functions, see, for example~\cite{Byrd:1971bey,NIST:DLMF}.

\subsection{Chebyshev's Theorem}
\label{sub:C_theorem}

A key theorem which states under which conditions certain integrals contain special functions is Chebyshev's theorem of integration~\cite{Tchebichef1853,marchisotto1994invitation}.

\begin{theorem}{Chebyshev's theorem of integration.}
The integral 
\begin{equation}
    \int x^{p}\left(\alpha+\beta x^r\right)^q \dd x\,,\quad r\neq 0\,,\quad p,q,r\in\mathbb{Q}\,,
\end{equation}
admits a representation in terms of elementary functions \emph{if and only if} at least one of 
\begin{equation*}
    (p+1)/r\,, \quad q\,, \quad (p+1)/r+q\,,
\end{equation*}
    is an integer.
\end{theorem}
\noindent All the other cases require the use of special functions.

\subsection{Elliptic integrals}
\label{subsec:elliptic}
For a complete overview of elliptic integrals, see~\cite[Chapter 19]{NIST:DLMF}. The \emph{incomplete elliptic integral of the first kind} $F$ is defined as
\begin{equation}
    F\left[\phi,m\right]=\int_0^{\phi} \frac{\dd\theta}{\sqrt{1-m\sin^2\theta}}=\int_0^{\sin\phi} \frac{\dd t}{\sqrt{(1-t^2)(1-mt^2)}}\,. \label{eqn:elliptick1} 
\end{equation} 
The \emph{incomplete elliptic integral of the third kind} $\Pi$ is defined as
\begin{equation}
    \Pi\left[n,\phi,m\right]=\int_0^{\phi} \frac{1}{1-n\sin^2\theta}\frac{\dd\theta}{\sqrt{1-m\sin^2\theta}}=\int_0^{\sin\phi} \frac{1}{1-nt^2}\frac{\dd t}{\sqrt{(1-t^2)(1-m t^2)}}\,. \label{eqn:elliptick2}
\end{equation}
where in both cases $t=\sin\theta$. Note that both integrals are said to be \emph{complete} when $\phi=\pi/2$, i.e.\ when $\sin\phi=1$. 

The \emph{complete elliptic integral of the first kind}, $K$, is defined as $K(m)=F[\pi/2,m]$, and the \emph{complete elliptic integral of the third kind} $\Pi$ is defined as $\Pi[n,m]=\Pi[n,\pi/2,m]$.

\subsection{Inverse Weierstrass elliptic function}
\label{subsec:weierstrassP}
Let us define
\begin{equation}
    \wp^{-1}(y)\equiv u=\int_{y}^{\infty}\frac{\dd t}{\sqrt{4t^3-g_2t-g_3}}\,, \label{eqn:inversewp}
\end{equation}
then the inverse function $y=\wp\left(u,g_2,g_3\right)\equiv\wp(u)$ defines the \emph{Weierstrass elliptic function}, see \cite{Byrd:1971bey} or again~\cite[Chapter 23]{NIST:DLMF}. 

\subsection{Hypergeometric function}\label{subsec:hypergeom}

The \emph{generalised hypergeometric function} is defined by the power series
\begin{equation*}
    {}_pF_q(a_1,\ldots,a_p;b_1,\ldots,b_q;z)=\sum_{n=0}^\infty \frac{\left(a_1\right)_n\cdots\left(a_p\right)_n}{\left(b_1\right)_n\cdots\left(b_q\right)_n}\frac{z^n}{n!}\,,
\end{equation*}
where $a_j,b_k\in\mathbb{C}$. Here $\left(q\right)_n$ is the \emph{(rising) Pochhammer symbol}, which is given by
\begin{equation*}
    \left(q\right)_n=\begin{cases}
    1 & n=0\\
    q(q+1)\cdots(q+n-1) & n>0
    \end{cases}.
\end{equation*}
For a detailed overview, see~\cite[Chapter 15]{NIST:DLMF}.

\addcontentsline{toc}{section}{References}
\bibliographystyle{jhepmodstyle}
\bibliography{bibliography}

\providecommand{\href}[2]{#2}\begingroup\raggedright\begin{thebibliography}{10}

\bibitem{SupernovaSearchTeam:1998fmf}
{\bf Supernova Search Team} Collaboration, A.~G. Riess et~al., {\it
  {Observational evidence from supernovae for an accelerating universe and a
  cosmological constant}},  Astron.\ J. {\bf 116} (1998) 1009--1038,
  [\href{http://arxiv.org/abs/astro-ph/9805201}{{\tt astro-ph/9805201}}].

\bibitem{SupernovaCosmologyProject:1998vns}
{\bf Supernova Cosmology Project} Collaboration, S.~Perlmutter et~al., {\it
  {Measurements of $\Omega$ and $\Lambda$ from 42 High Redshift Supernovae}},
  Astrophys.\ J. {\bf 517} (1999) 565--586,
  [\href{http://arxiv.org/abs/astro-ph/9812133}{{\tt astro-ph/9812133}}].

\bibitem{Wright:1992tf}
E.~L. Wright et~al., {\it {Interpretation of the Cosmic Microwave Background
  radiation anisotropy detected by the COBE differential microwave
  radiometer}},  Astrophys.\ J.\ Lett. {\bf 396} (1992) L13--L18.

\bibitem{WMAP:2006bqn}
{\bf WMAP} Collaboration, D.~N. Spergel et~al., {\it {Wilkinson Microwave
  Anisotropy Probe (WMAP) three year results: implications for cosmology}},
  Astrophys.\ J.\ Suppl. {\bf 170} (2007) 377,
  [\href{http://arxiv.org/abs/astro-ph/0603449}{{\tt astro-ph/0603449}}].

\bibitem{Planck:2019evm}
{\bf Planck} Collaboration, Y.~Akrami et~al., {\it {Planck 2018 results.\ VII.\
  Isotropy and Statistics of the CMB}},  Astron.\ Astrophys. {\bf 641} (2020)
  A7, [\href{http://arxiv.org/abs/1906.02552}{{\tt arXiv:1906.02552}}].

\bibitem{Friedman:1922kd}
A.~Friedman, {\it {On the Curvature of space}},  Z.\ Phys. {\bf 10} (1922)
  377--386.

\bibitem{Lemaitre:1927zz}
G.~Lemaitre, {\it {A Homogeneous Universe of Constant Mass and Growing Radius
  Accounting for the Radial Velocity of Extragalactic Nebulae}},  Annales Soc.\
  Sci.\ Bruxelles A {\bf 47} (1927) 49--59.

\bibitem{Robertson:1935zz}
H.~P. Robertson, {\it {Kinematics and World-Structure}},  Astrophys.\ J. {\bf
  82} (1935) 284--301.

\bibitem{Robertson:1935jpx}
H.~P. Robertson, {\it {Kinematics and World-Structure.\ 2}},  Astrophys.\ J.
  {\bf 83} (1935) 187--201.

\bibitem{Robertson:1936zz}
H.~P. Robertson, {\it {Kinematics and World-Structure.\ 3}},  Astrophys.\ J.
  {\bf 83} (1936) 257--271.

\bibitem{Walker:1937qxv}
A.~G. Walker, {\it {On Milne's Theory of World-Structure}},  Proc.\ Lond.\
  Math.\ Soc.\ s {\bf 2--42} (1937), no.~1 90--127.

\bibitem{Lachieze-Rey:1995qrb}
M.~Lachieze-Rey and J.-P. Luminet, {\it {Cosmic topology}},  Phys.\ Rept. {\bf
  254} (1995) 135--214, [\href{http://arxiv.org/abs/gr-qc/9605010}{{\tt
  gr-qc/9605010}}].

\bibitem{Hawking:1973uf}
S.~W. Hawking and G.~F.~R. Ellis, {\em {The Large Scale Structure of
  Space-Time}}.
\newblock Cambridge Monographs on Mathematical Physics. Cambridge University
  Press, 2, 2023.

\bibitem{Griffiths:2009dfa}
J.~B. Griffiths and J.~Podolsky, {\em {Exact Space-Times in Einstein's General
  Relativity}}.
\newblock Cambridge Monographs on Mathematical Physics. Cambridge University
  Press, Cambridge, 2009.

\bibitem{Harrison:1967zz}
E.~R. Harrison, {\it {Classification of Uniform Cosmological Models}},  Mon.\
  Not.\ Roy.\ Astron.\ Soc. {\bf 137} (1967) 69--79.

\bibitem{Barrow:1995cfa}
J.~D. Barrow and M.~P. Dabrowski, {\it {Oscillating Universes}},  Mon.\ Not.\
  Roy.\ Astron.\ Soc. {\bf 275} (1995) 850--862.

\bibitem{Barrow:2004ad}
J.~D. Barrow, D.~Kimberly, and J.~Magueijo, {\it {Bouncing universes with
  varying constants}},  Class.\ Quant.\ Grav. {\bf 21} (2004) 4289--4296,
  [\href{http://arxiv.org/abs/astro-ph/0406369}{{\tt astro-ph/0406369}}].

\bibitem{Brandenberger:2016vhg}
R.~Brandenberger and P.~Peter, {\it {Bouncing Cosmologies: Progress and
  Problems}},  Found.\ Phys. {\bf 47} (2017), no.~6 797--850,
  [\href{http://arxiv.org/abs/1603.05834}{{\tt arXiv:1603.05834}}].

\bibitem{Barrow:2017zar}
J.~D. Barrow and C.~Ganguly, {\it {The Shape of Bouncing Universes}},  Int.\
  J.\ Mod.\ Phys.\ D {\bf 26} (2017), no.~12 1743016,
  [\href{http://arxiv.org/abs/1705.06647}{{\tt arXiv:1705.06647}}].

\bibitem{Barrow:1986cur}
J.~D. {Barrow}, G.~J. {Galloway}, and F.~J. {Tipler}, {\it {The closed-universe
  recollapse conjecture.}},  Mon.\ Not.\ Roy.\ Astron.\ Soc. {\bf 223} (Dec.,
  1986) 835--844.

\bibitem{Khoury:2001wf}
J.~Khoury, B.~A. Ovrut, P.~J. Steinhardt, and N.~Turok, {\it {The Ekpyrotic
  universe: Colliding branes and the origin of the hot big bang}},  Phys. Rev.
  D {\bf 64} (2001) 123522, [\href{http://arxiv.org/abs/hep-th/0103239}{{\tt
  hep-th/0103239}}].

\bibitem{Steinhardt:2004gk}
P.~J. Steinhardt and N.~Turok, {\it {The Cyclic model simplified}},  New
  Astron. Rev. {\bf 49} (2005) 43--57,
  [\href{http://arxiv.org/abs/astro-ph/0404480}{{\tt astro-ph/0404480}}].

\bibitem{Lehners:2008vx}
J.-L. Lehners, {\it {Ekpyrotic and Cyclic Cosmology}},  Phys. Rept. {\bf 465}
  (2008) 223--263, [\href{http://arxiv.org/abs/0806.1245}{{\tt
  arXiv:0806.1245}}].

\bibitem{Linde:2002ws}
A.~D. Linde, {\it {Inflationary theory versus ekpyrotic / cyclic scenario}},
  in {\em {Workshop on Conference on the Future of Theoretical Physics and
  Cosmology in Honor of Steven Hawking's 60th Birthday}}, pp.~801--838, 5,
  2002.
\newblock \href{http://arxiv.org/abs/hep-th/0205259}{{\tt hep-th/0205259}}.

\bibitem{Pavlovic:2020sei}
P.~Pavlovi\'c and M.~Sossich, {\it {Dynamic properties of cyclic cosmologies}},
   Phys. Rev. D {\bf 103} (2021), no.~2 023529,
  [\href{http://arxiv.org/abs/2009.03625}{{\tt arXiv:2009.03625}}].

\bibitem{Bohmer:2016ome}
C.~G. B\"ohmer, {\em {Introduction to General Relativity and Cosmology}}.
\newblock Essential Textbooks in Physics. World Scientific, 12, 2016.

\bibitem{Baumann:2022mni}
D.~Baumann, {\em {Cosmology}}.
\newblock Cambridge University Press, 7, 2022.

\bibitem{andrews1999special}
G.~E. Andrews, R.~Askey, R.~Roy, R.~Roy, and R.~Askey, {\em Special functions},
  vol.~71.
\newblock Cambridge University Press, 1999.

\bibitem{armitage2006elliptic}
J.~V. Armitage and W.~F. Eberlein, {\em Elliptic functions}, vol.~67.
\newblock Cambridge University Press, 2006.

\bibitem{Halburd2017special}
R.~Halburd, {\it Special functions},  in {\em LTCC Advanced Mathematics Series:
  Volume 6 Analysis and Mathematical Physics}, ch.~4, pp.~109--138.
\newblock World Scientific, 2017.

\bibitem{Edwards:1972}
D.~{Edwards}, {\it {Exact expressions for the properties of the
  zero-pressureFriedmann models}},  Mon.\ Not.\ Roy.\ Astron.\ Soc. {\bf 159}
  (Jan., 1972) 51.

\bibitem{Edwards:1973}
D.~{Edwards}, {\it {Exact Solutions for Friedmann Models with Radiation}},
  Astrophys.\ Space Sci. {\bf 24} (Oct., 1973) 563--575.

\bibitem{Coquereaux:1981ya}
R.~Coquereaux and A.~Grossmann, {\it {Analytic Discussion of Spatially Closed
  Friedmann Universes With Cosmological Constant and Radiation Pressure}},
  Annals Phys. {\bf 143} (1982) 296.

\bibitem{DAmbroise:2009ruz}
J.~D'Ambroise, {\it {Applications of Elliptic and Theta Functions to
  Friedmann-Robertson-Lemaitre-Walker Cosmology with Cosmological Constant}},
  in {\em {MSRI summer graduate workshop: A Window into Zeta and Modular
  Physics}}, 8, 2009.
\newblock \href{http://arxiv.org/abs/0908.2481}{{\tt arXiv:0908.2481}}.

\bibitem{Chen:2014fqa}
S.~Chen, G.~W. Gibbons, Y.~Li, and Y.~Yang, {\it {Friedmann's Equations in All
  Dimensions and Chebyshev's Theorem}},  JCAP {\bf 12} (2014) 035,
  [\href{http://arxiv.org/abs/1409.3352}{{\tt arXiv:1409.3352}}].

\bibitem{Faraoni:2021opj}
V.~Faraoni, S.~Jose, and S.~Dussault, {\it {Multi-fluid cosmology in Einstein
  gravity: analytical solutions}},  Gen.\ Rel.\ Grav. {\bf 53} (2021), no.~12
  109, [\href{http://arxiv.org/abs/2107.12488}{{\tt arXiv:2107.12488}}].

\bibitem{Pavlov:2021vfv}
A.~E. Pavlov, {\it {Friedmann Cosmology in Elliptic Functions}},  Grav.\
  Cosmol. {\bf 27} (2021), no.~4 403--408.

\bibitem{DiValentino:2019qzk}
E.~Di~Valentino, A.~Melchiorri, and J.~Silk, {\it {Planck evidence for a closed
  Universe and a possible crisis for cosmology}},  Nature Astron. {\bf 4}
  (2019), no.~2 196--203, [\href{http://arxiv.org/abs/1911.02087}{{\tt
  arXiv:1911.02087}}].

\bibitem{Handley:2019tkm}
W.~Handley, {\it {Curvature tension: evidence for a closed universe}},  Phys.\
  Rev.\ D {\bf 103} (2021), no.~4 L041301,
  [\href{http://arxiv.org/abs/1908.09139}{{\tt arXiv:1908.09139}}].

\bibitem{Vagnozzi:2020rcz}
S.~Vagnozzi, E.~Di~Valentino, S.~Gariazzo, A.~Melchiorri, O.~Mena, and J.~Silk,
  {\it {The galaxy power spectrum take on spatial curvature and cosmic
  concordance}},  Phys.\ Dark Univ. {\bf 33} (2021) 100851,
  [\href{http://arxiv.org/abs/2010.02230}{{\tt arXiv:2010.02230}}].

\bibitem{Vagnozzi:2020dfn}
S.~Vagnozzi, A.~Loeb, and M.~Moresco, {\it {Eppur \`e piatto? The Cosmic
  Chronometers Take on Spatial Curvature and Cosmic Concordance}},  Astrophys.\
  J. {\bf 908} (2021), no.~1 84, [\href{http://arxiv.org/abs/2011.11645}{{\tt
  arXiv:2011.11645}}].

\bibitem{DiValentino:2020hov}
E.~Di~Valentino, A.~Melchiorri, and J.~Silk, {\it {Investigating Cosmic
  Discordance}},  Astrophys.\ J.\ Lett. {\bf 908} (2021), no.~1 L9,
  [\href{http://arxiv.org/abs/2003.04935}{{\tt arXiv:2003.04935}}].

\bibitem{Dhawan:2021mel}
S.~Dhawan, J.~Alsing, and S.~Vagnozzi, {\it {Non-parametric spatial curvature
  inference using late-Universe cosmological probes}},  Mon.\ Not.\ Roy.\
  Astron.\ Soc. {\bf 506} (2021), no.~1 L1--L5,
  [\href{http://arxiv.org/abs/2104.02485}{{\tt arXiv:2104.02485}}].

\bibitem{Anselmi:2022uvj}
S.~Anselmi, M.~F. Carney, J.~T. Giblin, S.~Kumar, J.~B. Mertens, M.~O'Dwyer,
  G.~D. Starkman, and C.~Tian, {\it {What is flat \ensuremath{\Lambda}CDM, and
  may we choose it?}},  JCAP {\bf 02} (2023) 049,
  [\href{http://arxiv.org/abs/2207.06547}{{\tt arXiv:2207.06547}}].

\bibitem{Glanville:2022xes}
A.~Glanville, C.~Howlett, and T.~M. Davis, {\it {Full-shape galaxy power
  spectra and the curvature tension}},  Mon.\ Not.\ Roy.\ Astron.\ Soc. {\bf
  517} (2022), no.~2 3087--3100, [\href{http://arxiv.org/abs/2205.05892}{{\tt
  arXiv:2205.05892}}].

\bibitem{Semenaite:2022unt}
A.~Semenaite, A.~G. S\'anchez, A.~Pezzotta, J.~Hou, A.~Eggemeier, M.~Crocce,
  C.~Zhao, J.~R. Brownstein, G.~Rossi, and D.~P. Schneider, {\it {Beyond
  \textendash{} \ensuremath{\Lambda}CDM constraints from the full shape
  clustering measurements from BOSS and eBOSS}},  Mon.\ Not.\ Roy.\ Astron.\
  Soc. {\bf 521} (2023), no.~4 5013--5025,
  [\href{http://arxiv.org/abs/2210.07304}{{\tt arXiv:2210.07304}}].

\bibitem{Planck:2018vyg}
{\bf Planck} Collaboration, N.~Aghanim et~al., {\it {Planck 2018 results. VI.
  Cosmological parameters}},  Astron. Astrophys. {\bf 641} (2020) A6,
  [\href{http://arxiv.org/abs/1807.06209}{{\tt arXiv:1807.06209}}]. [Erratum:
  Astron.Astrophys. 652, C4 (2021)].

\bibitem{Chandrasekhar:1985kt}
S.~Chandrasekhar, {\em {The mathematical theory of black holes}}.
\newblock Oxford University Press, 1985.

\bibitem{Lammerzahl:2015qps}
C.~L\"ammerzahl and E.~Hackmann, {\it {Analytical Solutions for Geodesic
  Equation in Black Hole Spacetimes}},  Springer Proc.\ Phys. {\bf 170} (2016)
  43--51, [\href{http://arxiv.org/abs/1506.01572}{{\tt arXiv:1506.01572}}].

\bibitem{Cieslik:2022uki}
A.~Cie\'slik and P.~Mach, {\it {Revisiting timelike and null geodesics in the
  Schwarzschild spacetime: general expressions in terms of Weierstrass elliptic
  functions}},  Class.\ Quant.\ Grav. {\bf 39} (2022), no.~22 225003,
  [\href{http://arxiv.org/abs/2203.12401}{{\tt arXiv:2203.12401}}].

\bibitem{Boehmer:2024kdt}
C.~G. Boehmer, A.~d'Alfonso~del Sordo, and B.~Hartmann, {\it {Azimuthal
  geodesics in closed FLRW cosmologies}},
  \href{http://arxiv.org/abs/2401.04597}{{\tt arXiv:2401.04597}}.

\bibitem{Einstein:1917ce}
A.~Einstein, {\it {Cosmological Considerations in the General Theory of
  Relativity}},  Sitzungsber.\ Preuss.\ Akad.\ Wiss.\ Berlin (Math.\ Phys.\ )
  {\bf 1917} (1917) 142--152.

\bibitem{Copeland:2006wr}
E.~J. Copeland, M.~Sami, and S.~Tsujikawa, {\it {Dynamics of dark energy}},
  Int.\ J.\ Mod.\ Phys.\ D {\bf 15} (2006) 1753--1936,
  [\href{http://arxiv.org/abs/hep-th/0603057}{{\tt hep-th/0603057}}].

\bibitem{Misner:1973prb}
C.~W. Misner, K.~S. Thorne, and J.~A. Wheeler, {\em {Gravitation}}.
\newblock W. H. Freeman, San Francisco, 1973.

\bibitem{Fier:2021fbt}
J.~Fier, X.~Fang, B.~Li, S.~Mukohyama, A.~Wang, and T.~Zhu, {\it {Gravitational
  wave cosmology: High frequency approximation}},  Phys. Rev. D {\bf 103}
  (2021), no.~12 123021, [\href{http://arxiv.org/abs/2102.08968}{{\tt
  arXiv:2102.08968}}].

\bibitem{Byrd:1971bey}
P.~F. Byrd and M.~D. Friedman, {\em {Handbook of Elliptic Integrals for
  Engineers and Scientists}}, vol.~67 of {\em Grundlehren der mathematischen
  Wissenschaften}.
\newblock Springer, 1971.

\bibitem{NIST:DLMF}
``{\it NIST Digital Library of Mathematical Functions}.''
  \url{https://dlmf.nist.gov/}, Release 1.2.1 of 2024-06-15.
\newblock F.~W.~J. Olver, A.~B. {Olde Daalhuis}, D.~W. Lozier, B.~I. Schneider,
  R.~F. Boisvert, C.~W. Clark, B.~R. Miller, B.~V. Saunders, H.~S. Cohl, and
  M.~A. McClain, eds.

\bibitem{Tchebichef1853}
P.~Tchebichef, {\it Sur l'intégration des différentielles irrationnelles.},
  Journal de Mathématiques Pures et Appliquées (1853) 87--111.

\bibitem{marchisotto1994invitation}
E.~A. Marchisotto and G.-A. Zakeri, {\it An invitation to integration in finite
  terms},  The College Mathematics Journal {\bf 25} (1994), no.~4 295--308.

\end{thebibliography}\endgroup

\end{document}